\documentclass{elsart}
\usepackage{amsmath}
\usepackage{amssymb}
\usepackage{color,setspace,times}
\usepackage{amssymb,amsmath,graphicx,color,rotating,subfigure,url}
\usepackage{lineno}
\usepackage[square,sort&compress,comma]{natbib}

\journal{Physica A} 

\begin{document}

\begin{frontmatter}

\title{Multifractal analysis of Chinese stock volatilities based on partition function approach}
\author[BS,SS]{Zhi-Qiang Jiang},
\author[BS,SS,CES,RCSE]{Wei-Xing Zhou\corauthref{cor}}
\corauth[cor]{Corresponding author. Address: 130 Meilong Road,
School of Business, P.O. Box 114, East China University of Science
and Technology, Shanghai 200237, China, Phone: +86 21 64253634, Fax:
+86 21 64253152.}
\ead{wxzhou@ecust.edu.cn} %


\address[BS]{School of Business, East China University of Science and Technology, Shanghai 200237, China}
\address[SS]{School of Science, East China University of Science and Technology, Shanghai 200237, China}
\address[CES]{Research Center for Econophysics, East China University of Science and Technology, Shanghai 200237, China}
\address[RCSE]{Research Center of Systems Engineering, East China University of Science and Technology, Shanghai 200237, China}

\begin{abstract}
We have performed detailed multifractal analysis on the minutely
volatility of two indexes and 1139 stocks in the Chinese stock
markets based on the partition function approach. The partition
function $\chi_q(s)$ scales as a power law with respect to box size
$s$. The scaling exponents $\tau(q)$ form a nonlinear function of
$q$. Statistical tests based on bootstrapping show that the
extracted multifractal nature is significant at the 1\% significance
level. The individual securities can be well modeled by the
$p$-model in turbulence with $p = 0.40 \pm 0.02$. Based on the idea
of ensemble averaging (including quenched and annealed average), we
treat each stock exchange as a whole and confirm the existence of
multifractal nature in the Chinese stock markets.
\end{abstract}

\begin{keyword}
Econophysics, Multifractal analysis, Partition function approach,
Quenched average, Annealed average, Bootstrapping, Stock markets
\end{keyword}

\end{frontmatter}

\section{Introduction}


\label{s1:intro}

Since the pioneering work in the 1990's
\cite{Mantegna-Stanley-1995-Nature,Ghashghaie-Breymann-Peinke-Talkner-Dodge-1996-Nature,Mantegna-Stanley-1996-Nature},
there has been a vigorous continuing investigation aimed at
discovering remarkable similarities between financial markets and
turbulent flows. Multifractal analysis, which was initially
introduced to investigate the intermittent nature of turbulence
\cite{Mandelbrot-1974-JFM,McCauley-1990-PR,Frisch-1996}, has also
been extensively applied to various financial time series
\cite{Zhou-2007}.
Multifractality has been regarded as one of the most important
stylized facts in equity returns. Many different methods have been
applied to characterize the hidden multifractal behavior in finance,
for instance, the fluctuation scaling analysis
\cite{Eisler-Kertesz-Yook-Barabasi-2005-EPL,Eisler-Kertesz-2007-EPL,Jiang-Guo-Zhou-2007-EPJB},
the structure function (or height-height correlation function)
method
\cite{Ghashghaie-Breymann-Peinke-Talkner-Dodge-1996-Nature,Vandewalle-Ausloos-1998-EPJB,Ivanova-Ausloos-1999-EPJB,Schmitt-Schertzer-Lovejoy-1999-ASMDA,Schmitt-Schertzer-Lovejoy-2000-IJTAF,Calvet-Fisher-2002-RES,Ausloos-Ivanova-2002-CPC,Gorski-Drozdz-Speth-2002-PA,Alvarez-Ramirez-Cisneros-Ibarra-Valdez-Soriano-2002-PA,Balcilar-2003-EMFT,Lee-Lee-2005a-JKPS,Lee-Lee-Rikvold-2006-PA},
multiplier method \cite{Jiang-Zhou-2007-PA}, multifractal detrended
fluctuation analysis (MF-DFA)
\cite{Kantelhardt-Zschiegner-Bunde-Havlin-Bunde-Stanley-2002-PA,Matia-Ashkenazy-Stanley-2003-EPL,Kwapien-Oswiecimka-Drozdz-2005-PA,Lee-Lee-2005b-JKPS,Oswiecimka-Kwapien-Drozdz-2005-PA,Jiang-Ma-Cai-2007-PA,Lee-Lee-2007-PA,Lim-Kim-Lee-Kim-Lee-2007-PA},
partition function method
\cite{Sun-Chen-Wu-Yuan-2001-PA,Sun-Chen-Yuan-Wu-2001-PA,Ho-Lee-Wang-Chuang-2004-PA,Wei-Huang-2005-PA,Gu-Chen-Zhou-2007-EPJB,Du-Ning-2008-PA,Zhuang-Yuan-2008-PA,Wei-Wang-2008-PA,Zhou-2007-JMSC,Jiang-Zhou-2008a-PA},
wavelet transform approaches
\cite{Struzik-Siebes-2002-PA,Turiel-Perez-Vicente-2003-PA,Turiel-Perez-Vicente-2005-PA,Oswiecimka-Kwapien-Drozdz-Rak-2005-APPB},
to list a few.

To the best of our knowledge, the first application of partition
function approach was carried out by Sun {\em{et al.}} to analyze
the multifractal nature of minutely Hang Seng Index (HSI) data for
individual trading days from January 3, 1994 to May 28, 1997
\cite{Sun-Chen-Wu-Yuan-2001-PA}. Ho {\em{et al.}} also reported that
there exists multifractal feature in the daily Taiwan Stock Price
Index (TSPI) data from 1987 to 2002
\cite{Ho-Lee-Wang-Chuang-2004-PA}. Wei and Huang analyzed the 5-min
intraday data of the Shanghai Stock Exchange Composite (SSEC) index
for individual trading days from January 1999 to July 2001
\cite{Wei-Huang-2005-PA}. They found a different empirical result
from the HSI and constructed a new measure of market risk to predict
index fluctuations. A similar multifractal feature was reported in
the high-frequency data of the SSEC index at different timescales
and time periods \cite{Du-Ning-2008-PA,Zhuang-Yuan-2008-PA}. What is
more interesting is the claim that the observed multifractal
singularity spectrum has predictive power for price fluctuations and
can serve as a measure of market risk
\cite{Sun-Chen-Yuan-Wu-2001-PA,Wei-Huang-2005-PA,Wei-Wang-2008-PA}.

The partition function approach was introduced to characterize
singular measures
\cite{Halsey-Jensen-Kadanoff-Procaccia-Shraiman-1986-PRA}, where a
spectrum of local singularities exist. When the measure is
homogenous in singularity (thus not a multifractal), the scaling
exponent $\tau(q)$ of the $q$-order partition function is linear
with respect to $q$. In all the aforementioned studies, the index
{\em{per se}} is treated as being proportional to a singular
measure. Intuitively, there is no singularity in the index. This
argument is indeed confirmed by the very narrow width of the
extracted singularity spectrum $f(\alpha)$. More precisely, the
minimum singularity strength $\alpha_{\min}$ and the maximum
singularity strength $\alpha_{\max}$ are both close to 1.
Alternatively, the scaling exponent function $\tau(q)$ is linear
against $q$, which disapproves the presence of multifractality. Zhou
discussed these issues using 5-min SSEC data and performed extensive
statistical tests \cite{Zhou-2007-JMSC}. Jiang and Zhou further
clarified the situation using intraday high-frequency data for HSI,
SZSC (Shenzhen Stock Exchange Composite), S\&P500, and NASDAQ
\cite{Jiang-Zhou-2008a-PA}.

Having said this, we nevertheless figure that the idea is valuable
to apply partition function approach to the multifractal analysis of
stocks and indexes and to the possible application of multifractal
properties in market prediction and risk management. This, of
course, should be done using returns rather than stock prices or
indexes. In this work, we shall follow this line to perform detailed
multifractal analysis on Chinese stocks and indexes, which form a
huge database. In almost all previous studies concerning the
multifractal nature of financial markets, the investigation was
conducted based on a single financial time series for individual
stocks or indexes. In this paper, we propose to study many stocks as
an ensemble. The idea is that different stocks in a same market
share many common underlying mechanics. This consideration leads to
the assumption that different stocks are realizations of a same
stochastic process and we can perform ensemble averaging in the
multifractal analysis as an analogue of diffusion-limited
aggregation
\cite{Cates-Witten-1987-PRA,Halsey-Leibig-1992-PRA,Halsey-Honda-Duplantier-1996-JSP,Halsey-Duplantier-Honda-1997-PRL}.

This paper is organized as follows. In Section~\ref{s2:pi}, we
describe the data sets and perform preprocesses. The methods adopted
are explained in detail in Section \ref{s3:method}. Section
\ref{s4:rd} presents the results and discussions. Finally, Section
\ref{s5:con} concludes.

\section{Preliminary information}
\label{s2:pi}

\subsection{The date sets}

We use a nice high-frequency database in the Chinese A-share stock
markets from January 2004 to June 2006. The trading rules were not
changed during this time period. After eliminating those stocks that
have recording errors or less than 0.1 million data points, we are
left with 715 stocks listed on Shanghai Stock Exchange (SHSE) and
424 stocks traded on the Shenzhen Stock Exchange (SZSE) (1139 in
total). We also include two indexes, the Shanghai Stock Exchange
Composite Index (SSEC) and the Shenzhen Stock Exchange Composite
Index (SZSC). The average size for each stock is about 1.32 million
and there are totally more than 1.5 billion data points for all the
stocks.

All of the these indexes and stock prices are tick-by-tick data,
which were recorded based on the market quotes disposed to all
traders in every six to eight seconds. Each datum is time stamped to
the nearest second at which one transaction occurs. Hence, the
recording time interval of price series is uneven for each stock.
For convenience, denote the time sequence for each data set as $t_i$
and the corresponding price sequence as $p(t_i)$, where
$i=1,2,\cdots$. In literature, $i$ is also called the event time.

\subsection{Definition of volatility}

For each time series, we define the event-time return $r(t_i)$ over
one event step as follows
\begin{equation}
r(t_i) = \ln p(t_i) - \ln p(t_{i-1})~.
 \label{Eq:return}
\end{equation}
We then calculate the minutely volatilities as follows,
\begin{equation}
v(t) = \sum_{t - \Delta t < \tau \leqslant t} |r(\tau)|~,
 \label{Eq:volatility}
\end{equation}
where $\Delta t = 1$ min. On each trading day, the Chinese stock
markets contain opening call auction (9:15 a.m. to 9:25 a.m.),
cooling period (9:25 a.m. to 9:30 a.m.), and continuous double
auction (9:30 a.m. to 11:30 a.m. and 13:00 p.m. to 15:00 p.m.).
Since July 1, 2006, the Shenzhen Stock Exchange introduced closing
call auction (14:57 to 15:00)\footnote{During the time period of our
data sets, both exchanges did not have closing call auction.}. Our
analysis focuses on the data recorded during the continuous double
auction. In this way, there are 240 data points of the volatility
for each time series on a single trading day.

\section{Methodologies}
\label{s3:method}

\subsection{Partition function approach}
\label{s31:pf}

Denote the minutely volatility series as $\{v(t): t = 1, 2, \cdot
\cdot \cdot, T\}$. Then the series is covered by $N$ boxes with
equal size $s$, where $N = [T/s]$. On each box, we define a quantity
$u$ as follows,
\begin{equation}
u(n;s) = u\left([(n-1)s+1,ns]\right)= \sum_{t=1}^{s} v[(n-1)s+t]~,
 \label{Eq:u}
\end{equation}
where $[(n-1)s+1,ns]$ is the $n$-th box. The box sizes $s$ are
chosen such that $N = [T/s]=T/s$. The measure $\mu$ on each box is
constructed as follows,
\begin{equation}
\mu(n) = \frac{u(n;s)}{\sum_{m=1}^{N} u(m;s)}~,
 \label{Eq:mu}
\end{equation}
We then calculate the partition function $\chi_q$
\cite{Halsey-Jensen-Kadanoff-Procaccia-Shraiman-1986-PRA}
\begin{equation}
\chi_q(s) =  \sum_{n=1}^{N} [\mu(n)]^q~,
 \label{Eq:chi}
\end{equation}
and expect it to scale as
\begin{equation}
\chi_q(s) \sim s^{\tau(q)}~,
 \label{Eq:chi:q}
\end{equation}
where the exponent $\tau(q)$ is a scaling exponent function. The
local singularity exponent $\alpha$ of the measure $\mu$ and its
spectrum $f(\alpha)$ are related to $\tau(q)$ through a Legendre
transformation
\cite{Halsey-Jensen-Kadanoff-Procaccia-Shraiman-1986-PRA}
\begin{equation} \label{Eq:alphaf}
\left\{ \begin{aligned}
         \alpha &= {\rm{d}}\tau(q)/{\rm{d}}q \\
                  f(\alpha)&=q \alpha -\tau(q)
                  \end{aligned} \right.~.
                          \end{equation}
When $u(n;s)/\sum u(m;s) \ll 1$ and $q \gg 1$, the estimation of the
partition function $\chi$ will be very difficult since the value is
so small that the computer is ``out of the memory.'' To overcome
this problem, we can calculate the logarithm of the partition
function, $\ln \chi_q(s)$, rather than the partition function
itself. A simple manipulation of Eqs.~(\ref{Eq:mu}) and
(\ref{Eq:chi}) results in the following formula,
\begin{equation}
\ln \chi_q(s) = \ln \sum_{n=1}^{N} \left[
\frac{u(n;s)}{\max\limits_m\{u(m;s)\}}\right]^q + q \ln \left[
\frac{\max\limits_m\{u(m;s)\}}{\sum_{m=1}^N u(m;s)}\right]~,
 \label{Eq:logchi}
\end{equation}
where $\max\limits_m\{u(m;s)\}$ is the maximum of $u(m;s)$ for $m=1,
2, \cdots, N$.

\subsection{Bootstrapping for statistical test}
\label{s31:boorstrap}

To test for the possibility that the empirical multifractality could
be artifactual, we adopt the following bootstrapping approach
\cite{Zhou-2007-JMSC,Jiang-Zhou-2008a-PA}. For a given volatility
time series, we reshuffle the series to remove any potential
temporal correlation and carry out the same multifractal analysis as
for the original data. We impose a very strict null hypothesis to
investigate whether the singularity spectrum $f(\alpha)$ is wider
than those produced by chance. The null hypothesis is the following:
\begin{equation}
~~~~H_0^1: \Delta \alpha \leqslant \Delta \alpha_{\rm{rnd}}~.
\end{equation}
The associated probability of false alarm for multifractality
(so-called ``false positive'' or error of type II) is defined by
\begin{equation}
p_1 = \frac{ \# [\Delta \alpha \leqslant \Delta
\alpha_{\rm{rnd}}]}{n},
 \label{Eq:p1}
\end{equation}
where $n$ is the number of shuffling and $\# [\Delta \alpha
\leqslant \Delta \alpha_{\rm{rnd}}]$ counts the number of $\Delta
\alpha$ whose value is not smaller than $\Delta \alpha_{\rm{rnd}}$.
As $n \rightarrow \infty$, it is clear that the estimated bootstrap
$p$-value will tend to the ideal bootstrap $p$-value. Under the
conventional significance level of $0.01$, the multifractal
phenomenon is statistically significant if and only if $p_1
\leqslant 0.01$. While $p_1 > 0.01$, the null hypothesis cannot be
rejected.

In a similar way, defining $F = [f(\alpha_{\min}) +
f(\alpha_{\max})]/2$, an analogous null hypothesis can be described
as follows:
\begin{equation}
H_0^2: F \geqslant F_{\rm{rnd}}~,
\end{equation}
where the false probability is
\begin{equation}
p_2 = \frac { \# [F \geqslant F_{\rm{rnd}}]}{n}~.
\end{equation}
Using the significance level of $0.01$, the multifractal phenomenon
is statistically significant if and only if $p \leqslant 0.01$.

\subsection{Ensemble average}
\label{s33:average}

In previous studies concerning multifractality in financial markets,
the multifractal analyses were carried out on individual stocks or
index series. The index can be regarded as an ensemble average of
the stock market in some sense. Here, we introduce a method to
investigate the multifractality in an ensemble of many stocks, which
is borrowed from the concept of computing multifractal dimensions in
Diffusion-Limited Aggregation
\cite{Cates-Witten-1987-PRA,Halsey-Leibig-1992-PRA,Halsey-Honda-Duplantier-1996-JSP,Halsey-Duplantier-Honda-1997-PRL}.
Regarding the stock market as a stochastic process, a stock trading
on the market can be considered as a realization of the stochastic
process. We define quenched and annealed mass exponent
$\tau_\mathcal{Q,A} (q)$ as follows
\begin{align}
\left\langle \ln \chi_{q}(s) \right\rangle &= -\tau_\mathcal {Q}(q) \ln s, \label{Eq:quenched}\\
\ln \left\langle \chi_{q}(s) \right\rangle &= -\tau_\mathcal {A}(q)
\ln  s, \label{Eq:Annealed}
\end{align}
where the angular brackets $\langle \cdot \rangle$ is the ensemble
average over all the chosen stocks. Intuitively, the annealed
exponents are more sensitive to rare samples of the ensemble with
unusual values of $\chi_q(s)$, while the quenched exponents are more
characteristic of typical members of the ensemble
\cite{Halsey-Duplantier-Honda-1997-PRL}.

\section{Results and discussions}
\label{s4:rd}

\subsection{Multifractal analysis}

\begin{figure}[htb]
\centering
\includegraphics[width=6.5cm]{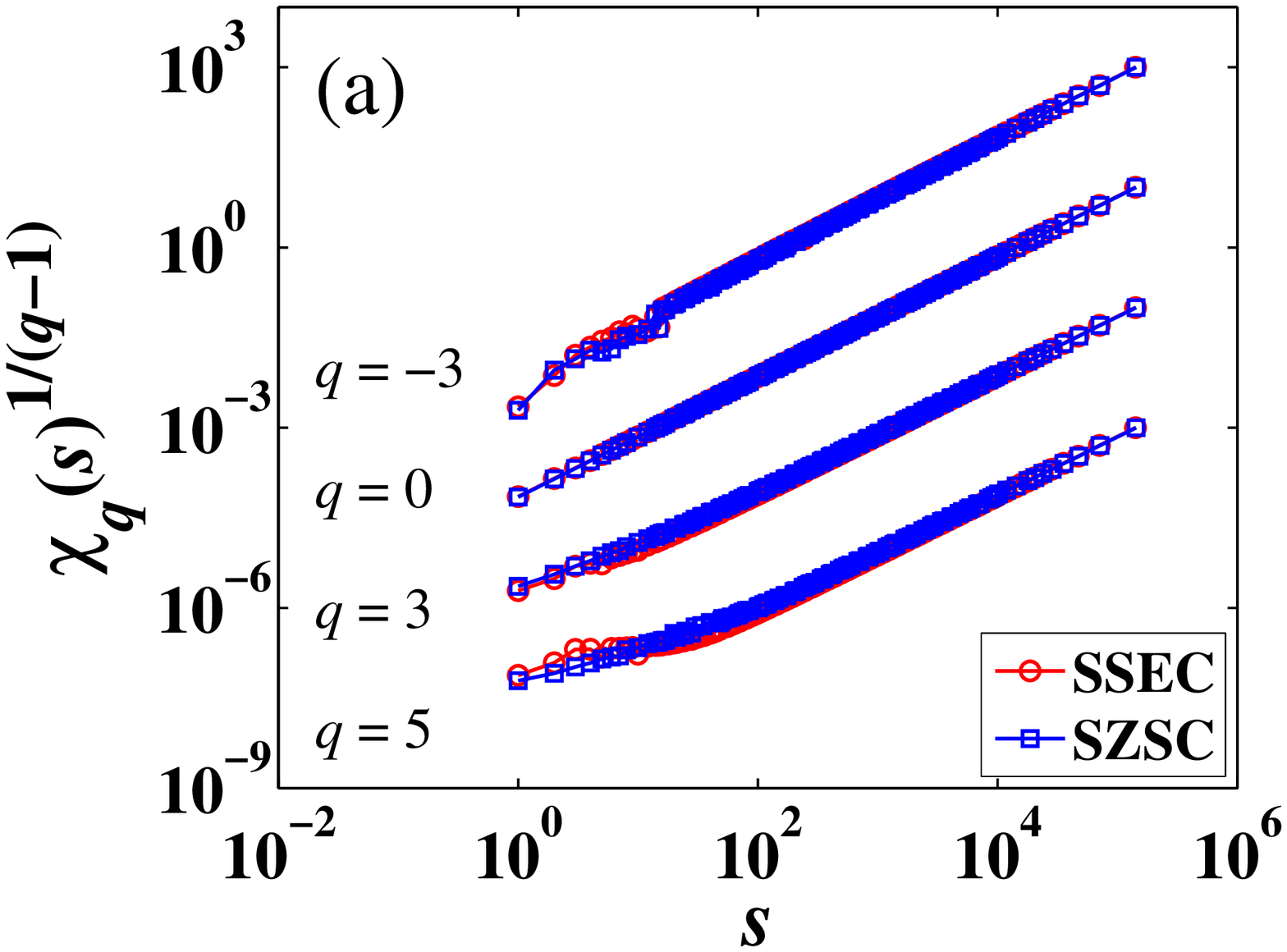}
\includegraphics[width=6.5cm]{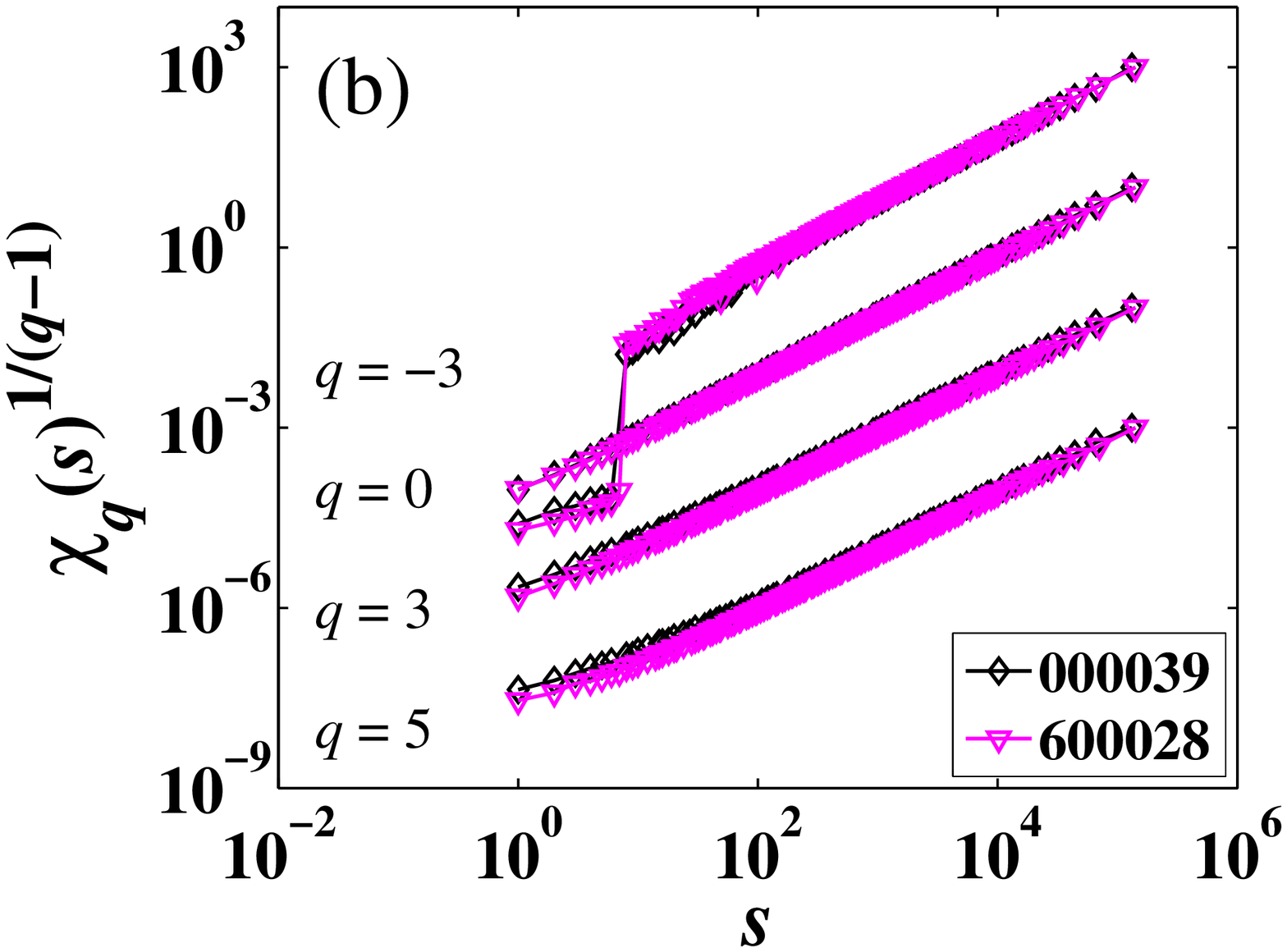}
\includegraphics[width=6.5cm]{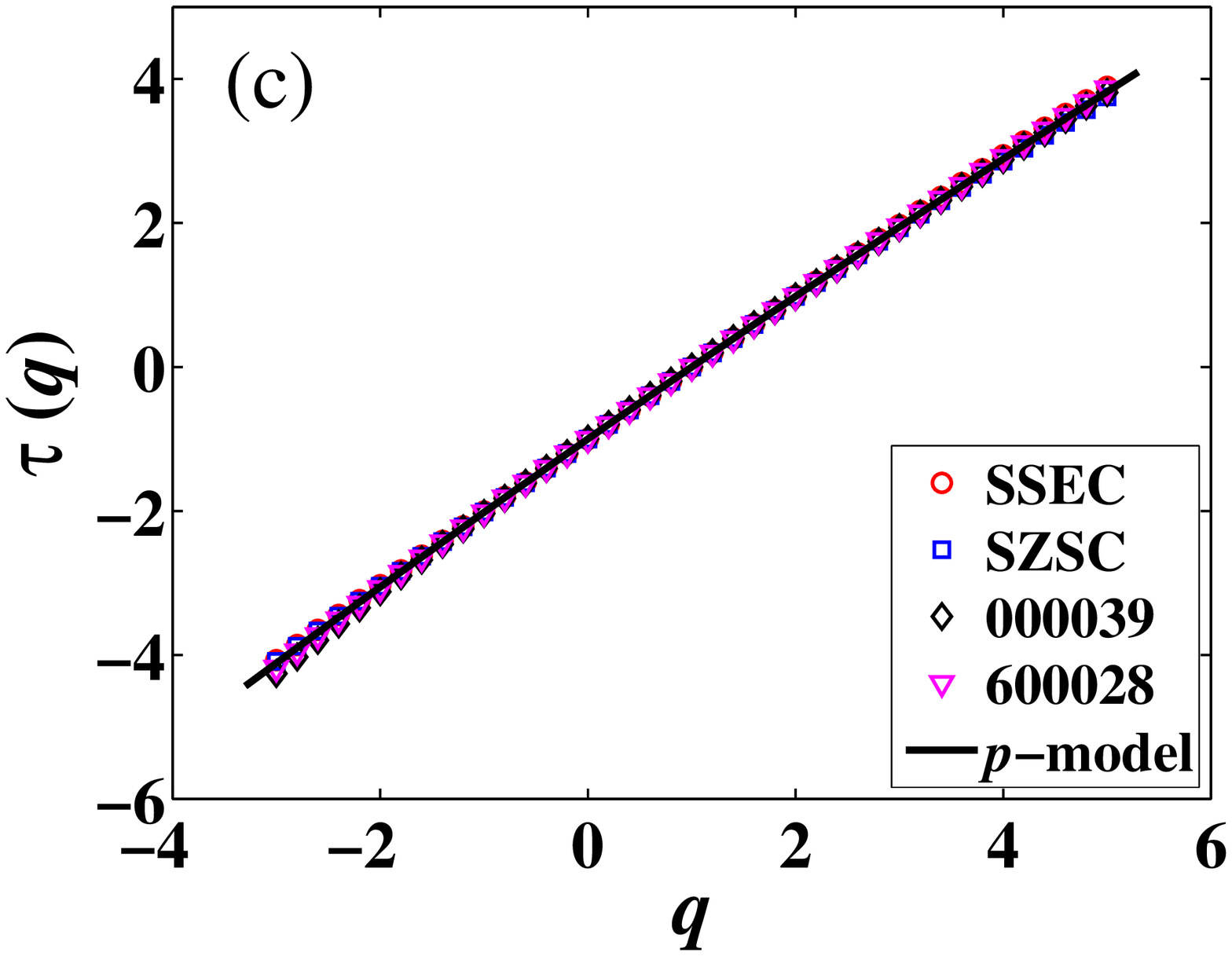}
\includegraphics[width=6.5cm]{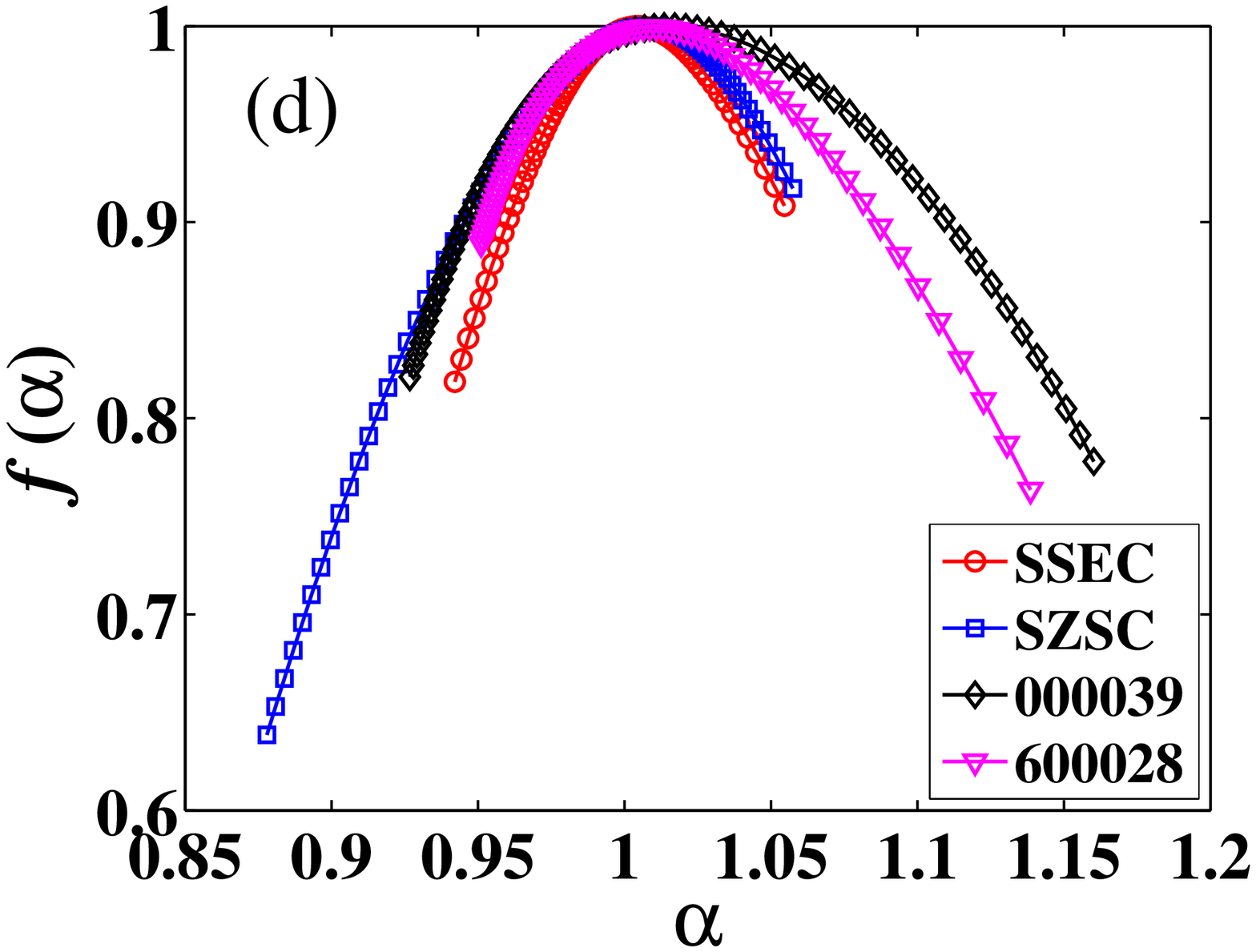}
\caption{(color online) Multifractal analysis. (a) Plots of
$\chi_q(s)^{1/(q-1)}$ as a function of box size $s$ for $q = -3, 0,
3$, and $5$ in log-log coordinates for two indexes. (b) Plots of
$\chi_q(s)^{1/(q-1)}$ as a function of box size $s$ for $q = -3, 0,
3$, and $5$ in log-log coordinates for two individual stocks. (c)
Dependence of scaling exponents $\tau(q)$ on the order $q$. (d)
Multifractal spectra $f(\alpha)$ obtained by Legendre transform of
$\tau(q)$.} \label{Fig:MF}
\end{figure}

Two important stock indexes (SSEC and SZSC) and two stocks (Sinopec,
600028, in Shanghai Stock Exchange and China International Marine
Containers, 000039, in Shenzhen Stock Exchange) are chosen as
examples to show multifractality in single index or stock volatility
series. Fig.~\ref{Fig:MF}(a) and (b) show the dependence of
$\chi_q(s)^{1/(q-1)}$ on box size $s$ for different values of $q$ in
log-log coordinates. Power laws with good quality are observed
between $\chi_q(s)^{1/(q-1)}$ and $s$. We also find that the scaling
range is wide and spans about three orders of magnitude. More
interestingly, when $s$ is small, there is a sudden jump on the
$\chi_{q}(s)^{1/(q-1)}$ curve for $q = -3$ in Fig.~\ref{Fig:MF}(b).
This is not surprising for individual stocks since they may have
time intervals within which the prices do not change. In this case,
there is at least one box with very vanishing measure $\mu$ when the
box size $s$ is smaller than or equal to some critical value $s_c$.
When $s>s_c$, the relative difference among measures is not large.
Hence, for all negative $q$, we will observe such jumps. This
explains the occurrence of a jump in individual stocks but not in
indexes. Consequently, in the determination of the scaling ranges,
we identify and exclude those jumps.

There are at least three cases corresponding to ``freezing'' price
in certain time period. The first case corresponds to the situation
that the price has reached its daily price limit. According to the
trading rules of SZSE and SHSE, there is a price fluctuation limit
of $\pm10\%$ for normal stocks and $\pm5\%$ for specially treated
(ST) stocks compared to the closing price of the last trading day.
When the price reach its daily price limit with a huge number of
shares waiting on the corresponding best bid or ask price, the price
does not change for a long time, which might last till the closure
of the market. The second case corresponds to those very liquid
stocks for which both the buy and sell sides of the order book are
very thick so that the price does not change frequently for most
marketable orders. The third case corresponds to those stocks with
very low transaction activities.

The scaling exponents $\tau(q)$ are given by the slopes of the
linear fits to $\ln \chi_q(s)$ with respect to $\ln s$ for different
values of $q$. We do this for all moments between -3 and 5 with an
increment of 0.2. Fig.~\ref{Fig:MF}(c) plots the dependence of the
mass exponents $\tau(q)$ on the moment order $q$.
Fig.~\ref{Fig:MF}(d) presents the multifractal singularity spectra
$f(\alpha)$ obtained through Legendre transformation of $\tau(q)$
defined by Eq.~(\ref{Eq:alphaf}). It is well-known that $\Delta
\alpha \triangleq \alpha_{\max} - \alpha_{\min}$ is an important
parameter qualifying the width of the extracted multifractal
spectrum. The larger is the $\Delta \alpha$, the stronger is the
multifractality. At the first glance, we find that $\Delta
\alpha_{\rm SSEC} < \Delta \alpha_{600028}$ and $ \Delta \alpha_{\rm
SZSC} < \Delta \alpha_{000039}$. This is in agreement with our
common sense that the fluctuations of index are less volatile than
that of individual stocks.

\begin{figure}[htb]
\centering
\includegraphics[width=8cm]{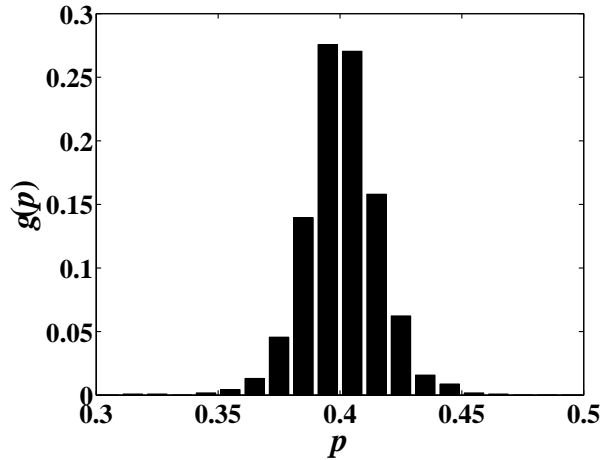}
\caption{Empirical occurrence frequency $g(p)$ of $p$.}
\label{Fig:P}
\end{figure}

To exhibit the analogue between financial market and fluid
mechanics, the $p$-model, which is a simple cascade model of energy
dissipation in fully developed turbulence
\cite{Meneveau-Sreenivasan-1987-PRL}, is applied to fit the mass
exponent functions and it gives an excellent parametrization of the
data. The theoretical mass function of the $p$-model can be
expressed as follows
\begin{equation}
 \tau(q) = -\frac{\ln[p^q+(1-p)^q]}{\ln 2}~.
\label{Eq:pmodel:tau}
\end{equation}
The solid line shown in Fig.~\ref{Fig:MF}(c) is drawn according to
the average of parameters obtained from fitting the four samples. We
see that the agreement between the $p$-model and the stock data is
remarkable for both positive and negative parts, indicating the
existence of information cascade process in stock market
\cite{Arneodo-Muzy-Sornette-1998-EPJB,Muzy-Delour-Bacry-2000-EPJB}.
We further extract the parameter $p$ of the $p$-model for all the
stocks under investigation. Fig.~\ref{Fig:P} illustrates that the
empirical occurrence frequency $g(p)$ as a function of $p$. We find
that $\langle{p}\rangle = 0.40 \pm 0.02$. In contrast, fully
developed turbulence gives $p = 0.3$
\cite{Meneveau-Sreenivasan-1987-PRL}.

\subsection{Statistical tests}

\begin{figure}[htb]
\centering
\includegraphics[width=6.5cm]{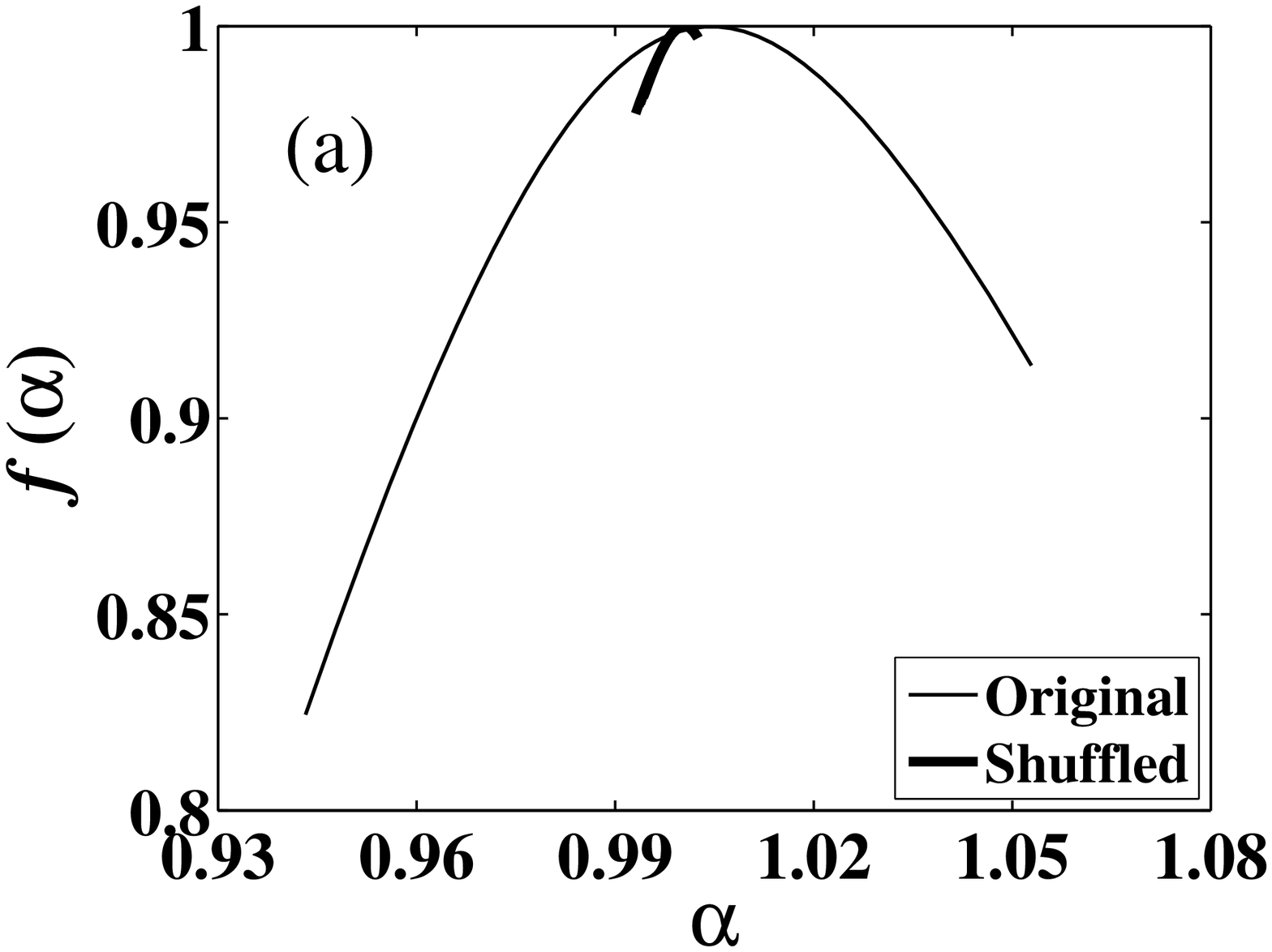}
\includegraphics[width=6.5cm]{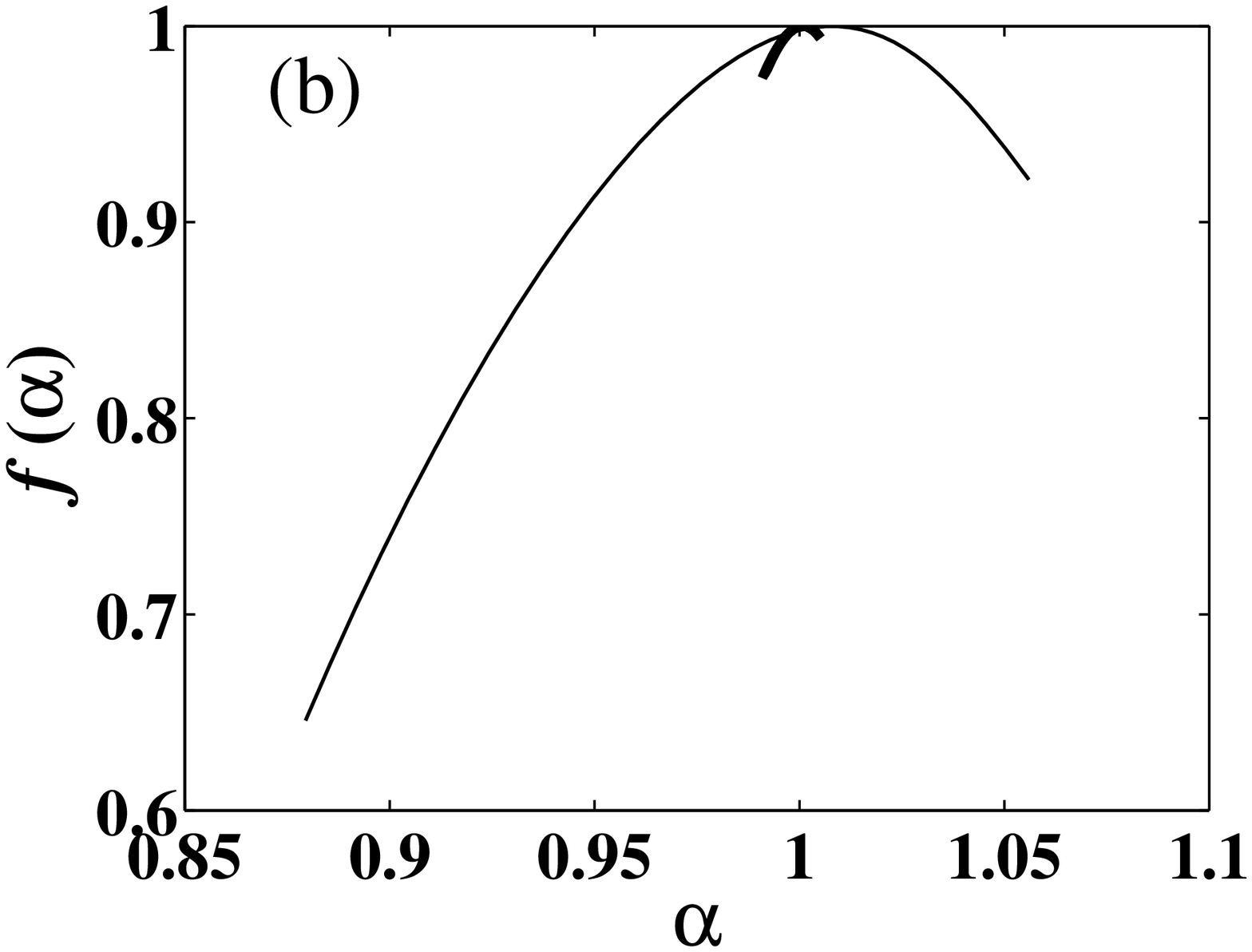}
\includegraphics[width=6.5cm]{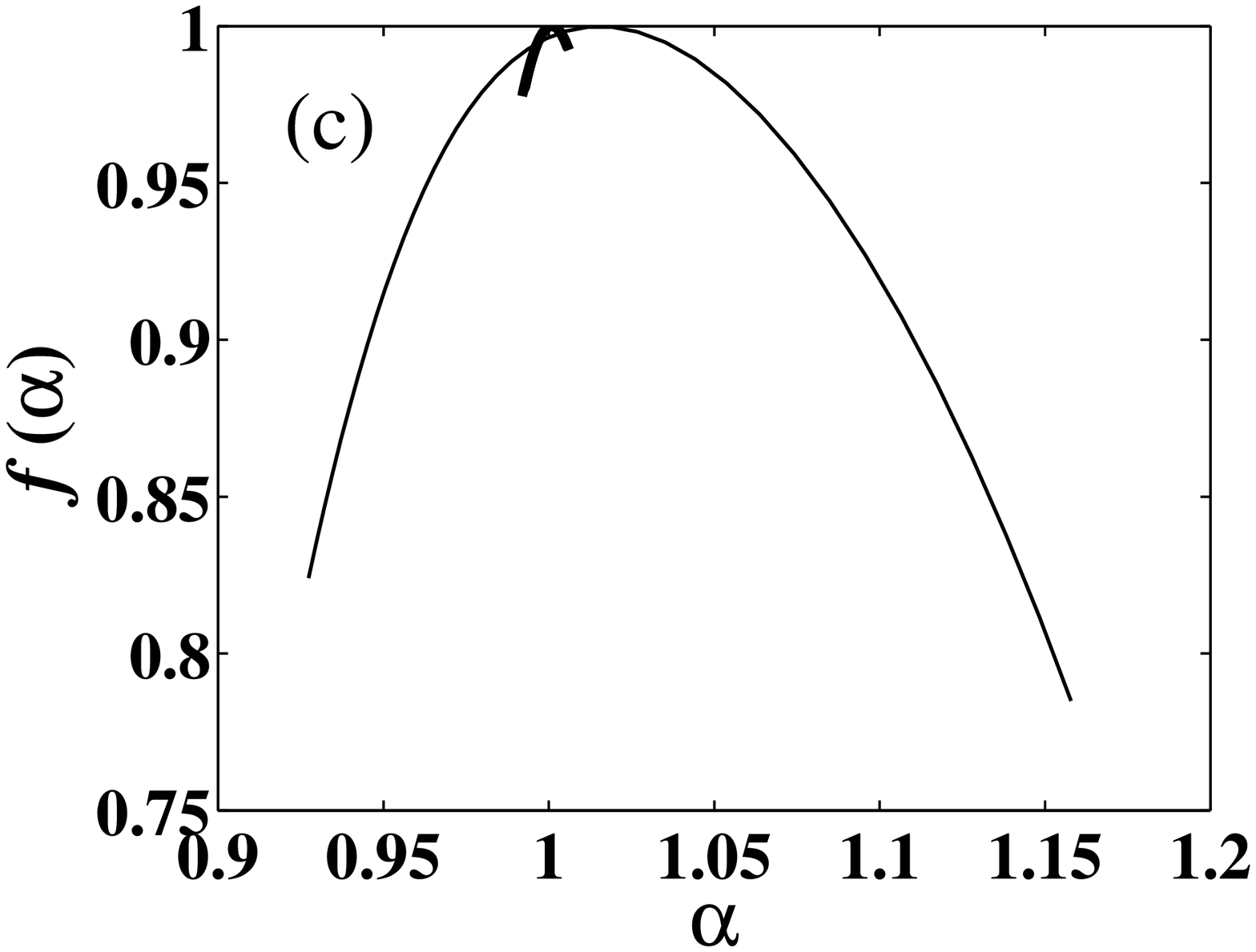}
\includegraphics[width=6.5cm]{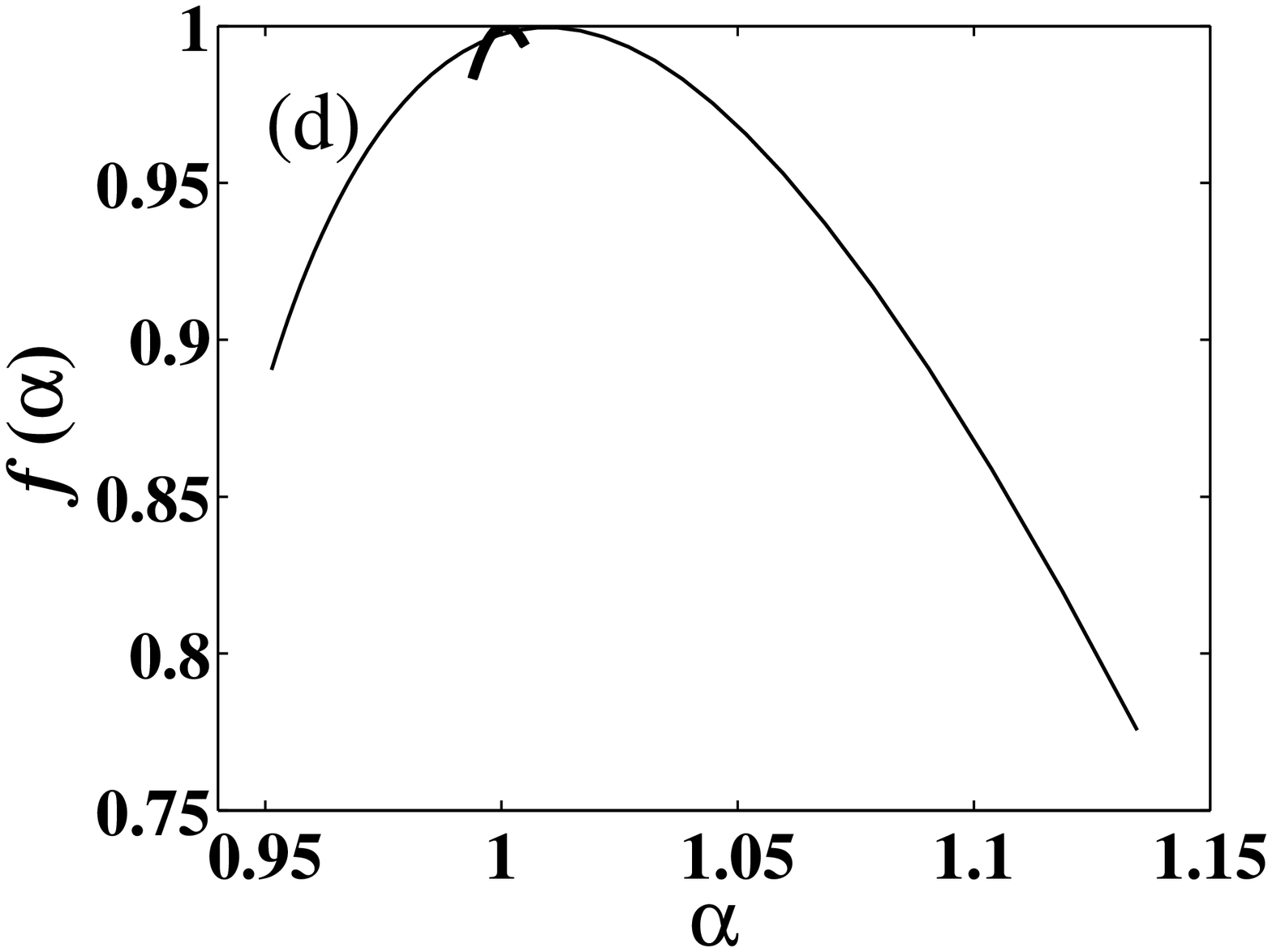}
\caption{Comparison between multifractal spectra extracted from real
and shuffled data. The thin lines correspond to the real data, while
the thick lines are for the shuffled data. (a) SSEC, (b) SZSC, (c)
000039, and (d) 600028.} \label{Fig:SFalpha}
\end{figure}

We assess the statistical significance of the empirical
multifractality in the sprit of bootstrapping tests. We reshuffle
the time series and perform the same multifractal analysis.
Fig.~\ref{Fig:SFalpha} compares the multifractal spectra of the raw
time series and that of the 10 shuffled time series for the two
chosen indexes and two securities. The thin lines are associated
with the real data, while the thick lines are obtained from the
shuffled data. An eye inspection already shows the deviation of the
multifractal spectra of the real data $f(\alpha)$ and that of the
shuffled data $f_{\rm{rnd}}(\alpha_{\rm{rnd}})$. We can infer that
one of the most important causations for the existence of the
multifractality is the long memory in the volatility series.

In the following, we shall give a more systemic statistical test on
Chinese stock market. For all the chosen series, we shuffle the data
1000 times for each stock and reinvestigate the multifractality of
the surrogate data. And the corresponding multifractal spectra are
obtained. For each singularity spectrum, we calculate two
characteristic quantities, $\Delta\alpha$ and $F \triangleq [
f(\alpha_{\min}) + f(\alpha_{\max}) ] / 2$. We find that $\Delta
\alpha > \langle \Delta \alpha_{\rm{rnd}} \rangle$ and $F <\langle
F_{\rm{rnd}} \rangle$ for all the series, which implies that there
are discrepancies between the multifractal spectra of the shuffled
data and that of the real data. We find that $p_1 = 0$ and $p_2 = 0$
for the SSEC and $p_1 = 0$ and $p_2 = 0$ for the SZSC, which
provides undoubtable evidence for the presence of multifractality in
the SSEC and SZSC data. Under the significance level of $1\%$, we
find that the multifractal nature is significant for all stocks.

\subsection{Quenched and annealed average}

\begin{figure}[htb]
\centering
\includegraphics[width=6.5cm]{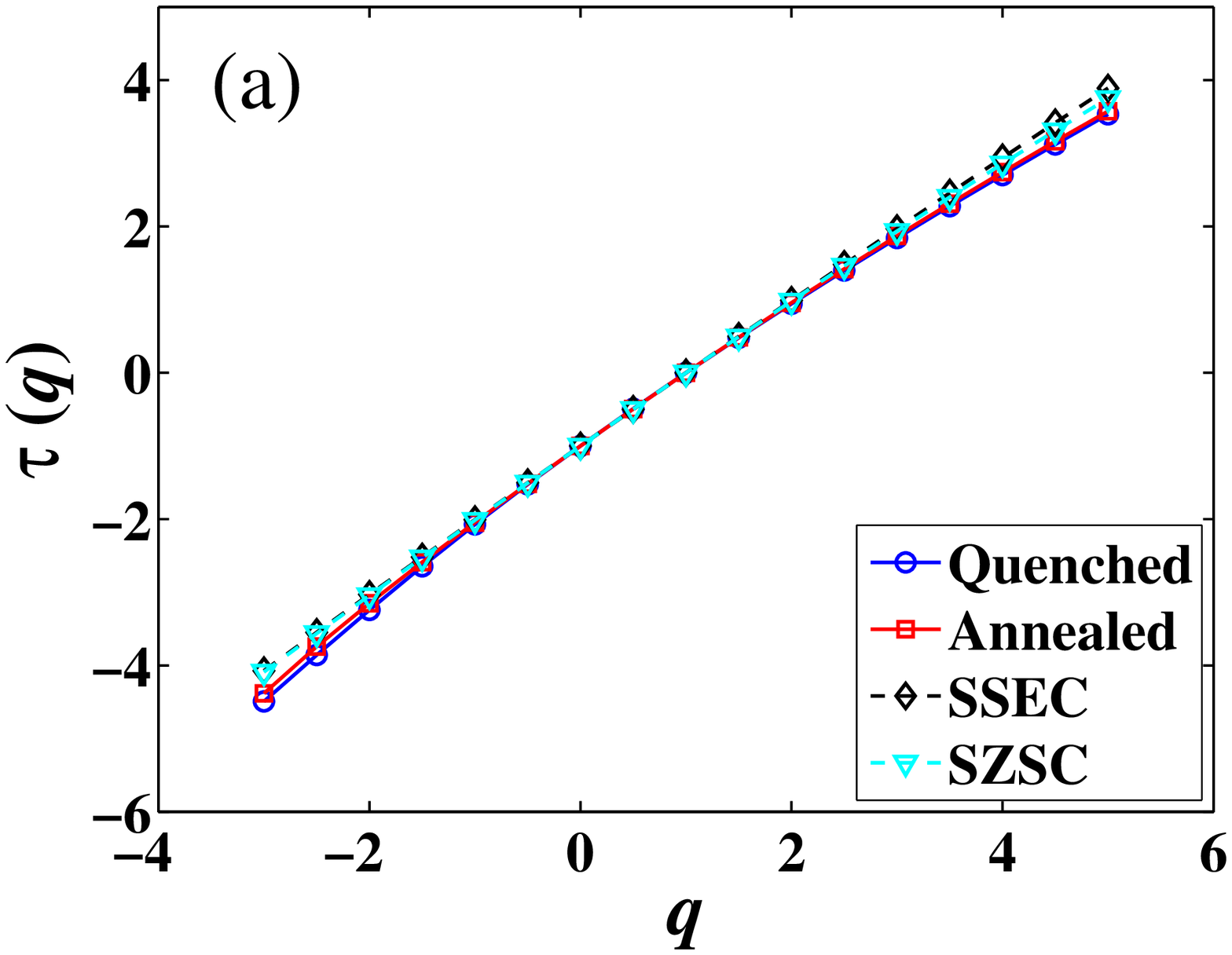}
\includegraphics[width=6.5cm]{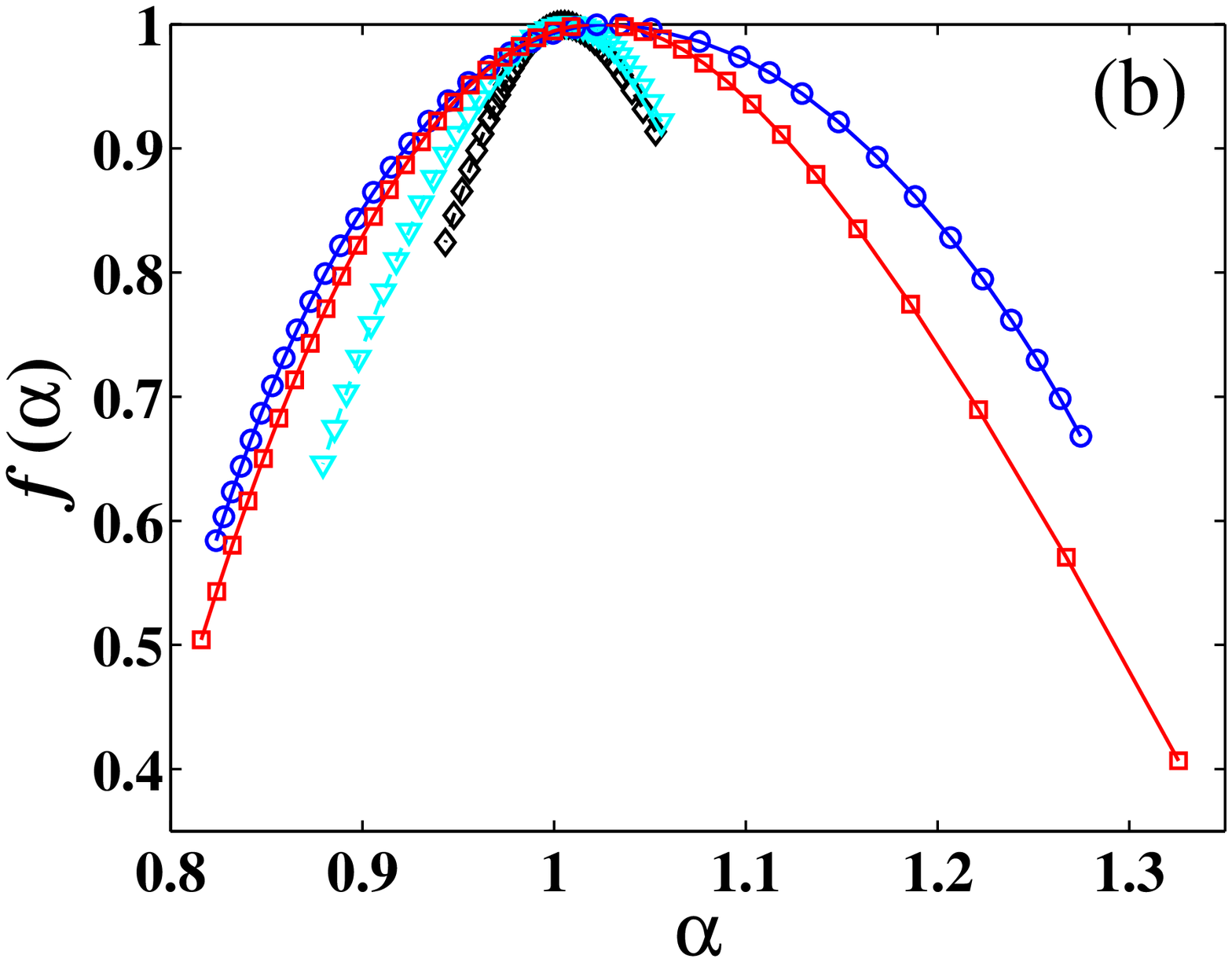}
\caption{(color online) Quenched and annealed multifractal analysis.
(a) Plots of mass exponents $\tau(q)$ with respect to the moment
order $q$. (b) Multifractal spectra.} \label{Fig:QA}
\end{figure}

Fig.~\ref{Fig:QA}(a) illustrates the quenched and annealed exponents
$\tau_\mathcal{Q,A}$ against moment order $q$. For comparison, we
also plot the mass exponents $\tau(q)$ of SSEC and SZSC. The
multifractal spectra obtained from Legendre transformation are
presented in Fig.~\ref{Fig:QA}(b). There are noticeable
discrepancies between $\Delta \alpha_\mathcal{A,Q}$ and $\Delta
\alpha_{\rm{SSEC,SZSC}}$. Indeed, ensemble averaging allows us to
capture the fluctuations among different realizations, which widens
the singularity spectrum when compared with that from individual
time series. This feature is well illustrated in Fig. 4(b).
Moreover, one can see that $\Delta \alpha_\mathcal{A}
> \Delta \alpha_\mathcal{Q}$. The appearance of this inequality
is linked to the sensitivity to rare values of the samples for
annealed average.

\section{Concluding remarks}
\label{s5:con}

We have performed detailed multifractal analysis on minutely
volatilities of two indexes and 1139 stocks in the Chinese stock
markets based on the partition function approach. The minutely
volatility is calculated as the sum of absolute returns in an
interval of one minute with higher-frequency data. A measure $\mu$
is constructed as the normalized volatility. Hence, the measure
$\mu$ is additive and conservational. This allows detailed
multifractal analysis based on the partition function approach. We
confirmed that $\mu$ is a multifractal measure.

According to our analysis, the partition function $\chi_q(s)$ for
each security scales as a power law with regard to box size $s$ for
each order $q$. The function of scaling exponents $\tau(q)$ for each
security can be estimated. The nonlinearity of $\tau(q)$ acts as a
hallmark of multifractality. The $p$-model in turbulence has been
used to fit the $\tau(q)$ function, resulting in a parameter $p =
0.40 \pm 0.02$. Statistical tests based on the bootstrapping
technique confirm the significance of the multifractal nature in the
volatility time series of individual securities.

An ensemble multifractal analysis was also carried out upon many
stocks and the annealed and quenched mass exponent functions
$\tau(q)$ have been determined. We note that the ensemble averaging
allows us to characterize the global multifractal properties of
large ensembles of data (the market as a whole) in a compact form
without dealing with details of individual realizations (individual
stocks). In this sense, we can draw a conclusion that the Chinese
stock market as a whole also exhibits multifractal behavior.

As discussed in Section \ref{s1:intro}, these correctly extracted
multifractal characteristics might have potential usefulness in
market prediction or risk management for individual securities or
the whole market. It is thus interesting to repeat the analysis in
literature
\cite{Sun-Chen-Yuan-Wu-2001-PA,Wei-Huang-2005-PA,Wei-Wang-2008-PA}
based on our results. However, this is beyond the scope of the
current work.

\bigskip
{\textbf{Acknowledgments:}}

We are indebted to Liang Guo for preprocessing the data. This work
was partly supported by the National Natural Science Foundation of
China (Grant No. 70501011), the Fok Ying Tong Education Foundation
(Grant No. 101086), the Shanghai Rising-Star Program (Grant No.
06QA14015), and the Program for New Century Excellent Talents in
University (Grant No. NCET-07-0288).

\bibliography{E:/Papers/Auxiliary/Bibliography} 

\end{document}